\documentclass[twocolumn,showpacs]{revtex4}

\usepackage{graphicx}
\usepackage{dcolumn}
\usepackage{bm}
\usepackage{epsfig}

\begin{document}

\title{Entanglement and decoherence of a micromechanical resonator
via coupling to a Cooper box}

\author{A. D. Armour,$^{1}$ M. P. Blencowe$^2$ and K. C. Schwab$^3$}
\affiliation{$^{1}$Blackett Laboratory, Imperial College of Science, 
Technology and Medicine, London SW7 2BW, United Kingdom\\
$^{2}$Department of Physics and Astronomy, Dartmouth College,
Hanover, New Hampshire 03755, USA\\
$^{3}$Laboratory for Physical Sciences,
College Park, Maryland 20740, USA}
\author{}
\affiliation{Department of Physics and Astronomy, Dartmouth College,
Hanover, New Hampshire 03755}

\begin{abstract}
We analyse the quantum dynamics of a micromechanical resonator 
capacitively coupled to  a Cooper box. 
With appropriate quantum state control of the Cooper box, 
the resonator can be driven into a superposition of 
spatially separated states. The Cooper box can also be used to 
probe the environmentally-induced decoherence of the resonator 
superposition state.  																																																																																																														
\end{abstract}

\maketitle

Micromechanical resonators with fundamental vibrational mode frequencies in 
the range 10MHz--1GHz can now be fabricated~\cite{cleland,carr}. 
Applications  include 
fast, ultra-sensitive force and displacement detectors~\cite{blencowe}, 
electrometers~\cite{cleland2,erbe}, and  
radio frequency signal processors~\cite{wang}.
However, the advances in the development of micromechanical devices also 
raise the fundamental question of whether mechanical systems containing
macroscopic numbers of atoms will exhibit quantum behavior.
Because of their size, quantum behavior in micromechanical systems will
be strongly influenced by interactions with the environment
and the existence of an experimentally accessible quantum regime will 
depend on the rate at which decoherence occurs~\cite{bose}.

In this Letter we analyse an experimentally implementable scheme  to  create 
and detect superpositions 
of macroscopically distinct quantum states  
in a micromechanical resonator, and furthermore measure their 
decoherence rates, by entangling the resonator with a Cooper 
box~\cite{bouchiat,nakamura1,makhlin}. The key 
advantage over optomechanical
schemes~\cite{bose,giovannetti} is the demonstrated coherent control
of the Cooper box quantum charge state~\cite{nakamura1}, together with the 
strong (controllable) coupling which can be achieved between the Cooper box
state and the motional degree of freedom of a micron-sized mechanical oscillator. 
Cooper box-based schemes have also been proposed  
for creating macroscopic quantum state superpositions in superconducting 
islands~\cite{marquardt} and  
superconducting resonators~\cite{buisson}. 

A Cooper box consists of a small superconducting island weakly-linked 
to a superconducting reservoir~\cite{bouchiat,nakamura1,makhlin}. 
The state of the Cooper box is 
determined by the balance between its Coulomb charging energy, and the 
strength of the Cooper-pair tunneling between the island and 
reservoir. Using an external gate, the Cooper box can be driven into 
either of two states of definite Cooper-pair number or a linear 
superposition of the two states~\cite{nakamura1}. Cooper boxes are 
being explored as possible candidates for qubits in future quantum computing 
devices since they act as readily controllable two-level quantum 
systems~\cite{makhlin,echternach}.

The electrostatic interaction between a  conducting
cantilever  and a nearby Cooper box causes a displacement in the cantilever
whose  sign depends on which of the two charge states the Cooper box is in.  
When the Cooper box is prepared in a superposition of charge states, 
it and the cantilever become entangled and the 
cantilever is driven into a superposition of spatially separated states.
If the coupling is strong enough, then the separation between the states in the 
superposition can become larger than their quantum position uncertainty, and
so we can describe them as macroscopically distinct.
Again using external voltage gates, the degree of entanglement 
between the cantilever 
and the Cooper box after a given period of interaction 
(which we call the wait time) can be
imprinted on the charge state of the box. For an isolated cantilever
the entanglement between the cantilever and the Cooper box is a periodic 
function of the wait time. However, because the cantilever is 
driven into a superposition of spatially separated states it will
be subject to environmental decoherence  which eventually destroys the 
periodicity in the entanglement between the cantilever and the Cooper box.
Of course the Cooper box itself is also subject to environmental decoherence, 
but this should not prevent the decoherence rate of the cantilever being 
determined (as we discuss below).

 The charge state
of the Cooper box can be measured with great sensitivity and with minimum 
disturbance using a radio-frequency single electron transistor
(rf-SET)~\cite{aassime}. Probing the 
charge state of the box after different wait times, and averaged over 
many different runs, will give information 
about the periodicity in the degree of entanglement 
of the cantilever and the Cooper box. Furthermore, measurement of the 
charge state of the Cooper box after different wait times  will also
allow the decoherence 
time of the cantilever due to interactions with its environment to be inferred. 
The circuit diagram for the system is shown in Fig.~1.

\begin{figure}[t]
\mbox{\epsfig{file=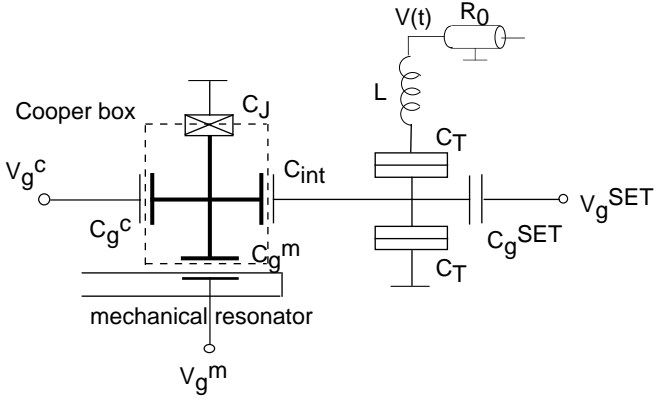, width=3.4in}}
\caption{Circuit diagram  for the coupled Cooper box-cantilever system and the
rf-SET.}
\end{figure}

Let us first focus on the dynamics of the Cooper box-coupled 
cantilever  system, neglecting 
the coupling to the cantilever environment and the rf-SET. 
The Hamiltonian  is
\[H=4 E_{\rm C} \delta n\hat{\sigma}_{z} -\frac{1}{2} 
E_{\rm J}\hat{\sigma}_{x} +\hbar\omega_{\rm m}\hat{a}^{\dagger}\hat{a} 
 +\lambda(\hat{a}+\hat{a}^{\dagger})\hat{\sigma}_{z},\]
where $\delta n=n_{\rm g}-(n+1/2)$ with 
$n_{\rm g}=-(C_{\rm g}^{\rm c} V_{\rm g}^{\rm c}+
C_{\rm g}^{\rm m}V_{\rm g}^{\rm m})/2e$ 
 the dimensionless, total gate charge. The control gate voltage  
$V_{\rm g}^{\rm c}$ and 
cantilever gate electrode voltage $V_{\rm g}^{\rm m}$ ranges are  
restricted such 
that $0\leq \delta n\leq 1/2$ for some chosen $n$, so that only Cooper charge 
states $\left|n\right.\rangle \equiv|-\rangle\equiv \left(_0^1 \right)$ and
$\left|n+1\right.\rangle\equiv|+\rangle \equiv \left(_1^0 \right)$ play a role. Thus 
it is natural to use spin notation where $\hat{\sigma}_{x}$ and 
$\hat{\sigma}_{z}$ are the usual Pauli matrices. The 
coupling constant between the box and cantilever electrode is 
$\lambda=-4E_{\rm C}n_{\rm g}^{\rm m}\frac{\Delta x_{\rm zp}}{d}$, 
where $n_{\rm g}^{\rm m}=-C_{\rm g}^{\rm m}V_{\rm g}^{\rm m}/2e$,  
$\Delta x_{\rm zp}$ is the zero-point displacement uncertainty of the 
cantilever, and $d\gg\Delta x_{\rm zp}$ 
is the cantilever electrode-island gap. Only the in-plane fundamental 
flexural mode  of the cantilever, with frequency $\omega_{\rm m}$  and
operators $a$ and $a^{\dagger}$, is taken into account. All other modes have 
a much weaker coupling to the box and will be neglected~\cite{armour}.  We 
assume that the Josephson junction capacitance $C_{\rm J}\gg C_{\rm 
g}^{\rm c}\ {\rm and}\ C_{\rm 
g}^{\rm m}$, so that the charging energy of the box 
$E_{\rm C}\approx {\rm e}^{2}/2C_{\rm J}$.

The scheme for the  control pulse sequence is indicated in Fig.~2. 
It is convenient to determine the evolution of the box-cantilever 
system using the coherent state basis for the cantilever. At 
$t=0$, we take as initial state 
$\left|\Psi_{0}\right.\rangle=
\left|-\right.\rangle\left|\alpha\right.\rangle$, 
where $\left|\alpha\right.\rangle$ denotes a coherent state~\cite{mandel}. The 
first pulse takes the box to the degeneracy point and is of duration 
$T_{\rm R}/4$, where $T_{\rm R}=h/E_{\rm J}$ is the coherent 
oscillation (Rabi) period of the Cooper state. The state 
$\left|\Psi_{0}\right.\rangle$ evolves to 
$|\Psi_{T_{\rm 
R}/4}\rangle=\frac{1}{\sqrt{2}}\left(\left|-\right.\rangle-
i\left|+\right.\rangle\right)\left|\alpha\right.\rangle$, where it 
is assumed that $\omega_{\rm m}T_{\rm R}\ll 1$ and the cantilever 
box coupling strength is such that the coherent state evolution can be 
neglected. Following the first pulse, there is a wait time, $\tau$, 
during which the box and cantilever systems interact, resulting in an 
entangled state: 
\begin{eqnarray*}
|\Psi_{T_{\rm 
R}/4+\tau}\rangle&=&\frac{1}{\sqrt{2}}{\rm e}^{2iE_{\rm C}\tau/\hbar}
\left|-\right.\rangle\left|\alpha_{-}(\tau)\right.\rangle\\
&&-\frac{i}{\sqrt{2}}{\rm e}^{-2i E_{\rm C}\tau/\hbar}\left|+\right.\rangle
\left|\alpha_{+}(\tau)\right\rangle,
\end{eqnarray*}
where we assume that $ E_{\rm J}\ll E_{\rm C}$, and so neglect the 
Josephson tunneling term in the evolution, and where
$\left|\alpha_{\pm}(\tau)\right\rangle={\rm e}^{\pm i\phi(\alpha,\tau)}
\left|\alpha {\rm e}^{-i\omega_{\rm 
m}\tau} \mp\kappa(1-{\rm e}^{-i\omega_{\rm 
m}\tau})\right\rangle,
$
with the phase $\phi(\alpha,\tau)=\frac{i\kappa}{2}[\alpha(1-{\rm e}^{-i\omega_{\rm 
m}\tau})-\alpha^{*}(1-{\rm e}^{i\omega_{\rm 
m}\tau})]$ and the dimensionless coupling $\kappa=\lambda/\hbar\omega_{\rm m}$.  
The spatial separation between the cantilever states 
$|\alpha_{\pm}(\tau)\rangle$ is ${2\kappa}
(1-\cos\omega_{\rm m}\tau)\Delta x_{\rm zp}$ and, thus, the 
condition for the  maximum separation of the states   to exceed their 
width is: $4|\kappa|>1$.

\begin{figure}[t]
\mbox{\epsfig{file=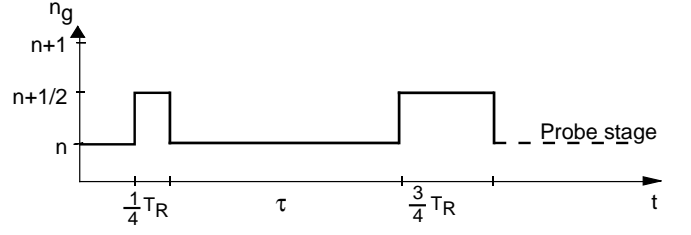, width=3.4in}}
\caption{Pulse sequence for manipulating the state of the box.}
\end{figure}

By taking the box to the 
degeneracy point a second time with a pulse of duration $3T_{\rm 
R}/4$, a    
signature of the separated cantilever states is imprinted on the Cooper pair number 
probabilities:
\begin{eqnarray*}|\Psi_{T_{\rm 
R}+\tau}\rangle&=& \frac{1}{2}|-\rangle\left[{\rm e}^{2i E_{\rm C}\tau/\hbar} 
|\alpha_{-}(\tau)\rangle+{\rm e}^{-2i E_{\rm C}\tau/\hbar} 
|\alpha_{+}(\tau)\rangle\right]\\
&-&\frac{i}{{2}}|+\rangle\left[{\rm e}^{2iE_{\rm C}\tau/\hbar} 
|\alpha_{-}(\tau)\rangle-{\rm e}^{-2iE_{\rm C}\tau/\hbar} 
|\alpha_{+}(\tau)\rangle\right]
\end{eqnarray*}
and 
\begin{equation}
{\rm P}(|-\rangle)=\frac{1}{2}\left\{1+\cos \left[4 E_{\rm C}\tau/\hbar 
+\phi(\alpha,\tau)\right]{\rm e}^{-4\kappa^2
(1-\cos\omega_{\rm m}\tau)}\right\}. \label{on}
\end{equation}
If there is no  coupling between the Cooper box and cantilever (i.e., 
$\kappa=0$), 
the second control pulse simply returns the box to its 
initial state $|-\rangle$ (the Cooper state has effectively performed a 
full Rabi oscillation at the degeneracy point) provided 
$\tau=2\pi k\hbar/4 E_{\rm 
C}$, $k=0,1,2\dots$.

Assuming that, before the control pulse sequence is applied, the 
box-cantilever system is in a thermal equilibrium state (because 
$4 E_{\rm C}\gg k_{\rm B}T$, the box will be in its ground state 
$|-\rangle$ to a good approximation),  we must thermally average the 
above probability. This gives
\begin{eqnarray}
{\rm P}_{\rm th}(|-\rangle)&=&
\frac{1}{2}\left\{1+\cos \left[4E_{\rm 
C}\tau/\hbar+4\kappa^2\sin\omega_{\rm m}\tau\right] \right.\nonumber \\
&&\left. \times
{\rm e}^{-4\kappa^{2}(1-\cos\omega_{\rm m}\tau)(1 +2\bar{N})}\right\}. \label{one}
\end{eqnarray}
where $\bar{N}=({\rm e}^{\hbar\omega_{\rm m}/k_{\rm B}T}-1)^{-1}$ is the 
thermal occupation of the cantilever mode. The 
 cosine function leads to rapid oscillations whose magnitude is 
controlled by the exponential term. It is convenient to define the 
envelope of ${\rm P}_{\rm th}(|-\rangle)$ as the function in Eq.\ (\ref{one})
with the argument in the square brackets set to zero.  

Notice that the envelope of Eq.\ (\ref{one}) recovers its initial value 
(i.e., unity)  as $\tau$
approaches the period $\tau_{\rm m}$ of the cantilever mode. 
This is a consequence of the harmonic nature of the cantilever as a 
measuring device for the Cooper box state; the correlations set up between the 
box  and cantilever states are completely undone and the two systems are 
no longer entangled after an integral number of harmonic oscillation periods. 
This `recoherence' effect is discussed in Ref.~\onlinecite{bose} for a 
system involving a cavity field coupled to a movable mirror. Similar
effects are also discussed in Refs.~\onlinecite{marquardt} and 
\onlinecite{buisson}.

The conditions for the quantum state control are as follows:
\[
\tau_{\rm j}<\frac{h}{4E_{\rm C}}\ll\frac{h}{E_{\rm 
J}}\ll\tau_{\rm m}<\tau^{\rm cb}_{\rm d},
\]
where $\tau_{\rm j}$ denotes the jitter time of the pulse 
sequence generator and $\tau^{\rm cb}_{\rm d}$ denotes 
the decoherence time of 
the Cooper box superposition states through processes other than due 
to the cantilever and its environment. The first inequality in the 
chain is necessary to resolve the rapid oscillations with period 
${h}/{4E_{\rm C}}$ in Eq.\ (\ref{one}), and thereby measure the associated 
envelope function; 
without being able to position the pulses with sufficient temporal
accuracy, the oscillations would be washed out giving a 
constant ${\rm P_{th}}(|-\rangle)\approx 1/2$. The last inequality is 
necessary to observe the recoherences and the effects of the 
cantilever's environment (which we discuss below). The middle two 
inequalities are not  essential, their purpose being only to simplify 
the theoretical analysis and hence the description of the 
quantum dynamics. A $1~{\rm ps}$ jitter time is 
achievable. Choosing $E_{\rm C}=150~\mu{\rm eV}$ gives $h/4 E_{\rm 
C}\approx 7~{\rm ps}$ and choosing $E_{\rm J}=4~\mu{\rm eV}$ gives 
$h/E_{\rm J}\approx 1~{\rm ns}$. A fundamental flexural frequency  
$\nu_{\rm m}=50~{\rm MHz}$, giving a period  $\tau_{\rm m}=20~{\rm 
ns}$, is 
readily achievable with micron-sized 
cantilevers~\cite{cleland,carr}.

The most serious practical constraint arises from the decoherence of the
Cooper-box itself, which if it occurs too fast will obscure the 
quantum dynamics
of the cantilever. At present, decoherence times of only a few ns have been 
achieved and an improvement of about an
order of magnitude would be required to implement our scheme. 
However, recent work by Nakamura {\it et al.}~\cite{nakamura2} has 
demonstrated that decoherence times of the box can be extended
by applying refocusing pulses. There is no fundamental reason
why the Cooper box decoherence time should be limited to less than 
$20~{\rm ns}$ and so considerable further improvements are to be expected.

In order that the Cooper-pair superposition state separate the 
cantilever coherent states by more than their width (the quantum position 
uncertainty), we require that 
the coupling strength satisfies $4|\lambda|/\hbar\omega_{\rm 
m}>1$. A Si cantilever with dimensions $l~({\rm length})\times w~({\rm 
width})\times t~({\rm thickness})= 1.6~\mu{\rm m}\times 0.1~\mu{\rm 
m}\times 0.1~\mu{\rm m}$ has a fundamental flexural frequency
$\nu_{\rm m}\approx 50~{\rm MHz}$ and zero-point uncertainty $\Delta 
x_{\rm zp}\approx 1.4\times 10^{-3}~{\rm \AA}$. Assuming a cantilever 
electrode-Cooper island gap $d=0.1~\mu{\rm m}$ and gate capacitance 
$C_{\rm g}^{\rm m}\approx 20~{\rm aF}$, the dimensionless gate
charge $n_{\rm g}^{\rm m}\approx -63V_{\rm g}^{\rm m}$. Substituting in these 
parameter values and  $E_{\rm C}=150~\mu{\rm 
eV}$, we have for the separation condition: $V_{\rm g}^{\rm m}> 
1.0~{\rm V}$. Such a voltage can be applied across a 
$0.1~\mu{m}$ gap: it  will 
deflect the cantilever by a much smaller distance 
than the gap and is well below the breakdown voltage.

We now turn to consider the effect of the cantilever's environment on
the coupled Cooper box-cantilever dynamics. We model the environment of the 
cantilever
as a bath of oscillators at a fixed temperature, $T$, each of which are weakly 
coupled to the fundamental flexural mode. This model is widely used 
for open systems and 
is equivalent to treating the cantilever mode as a damped quantum 
oscillator~\cite{mandel,armour},
characterised by an energy damping rate parameter, $2\gamma\ll 
\omega_{\rm m},k_{\rm B} T/\hbar$~\cite{valid}. When the 
calculation of ${\rm P_{th}}(|-\rangle)$ is repeated including the 
coupling of the cantilever to the bath oscillators we find
\begin{equation}
{\rm P}_{\rm th}(|-\rangle)=
\frac{1}{2}\left\{1+\cos \left[4E_{\rm 
C}\tau/\hbar+4\kappa^2\varphi(\tau)\right] 
{\rm e}^{-\Gamma(\tau)}\right\}, \label{two}
\end{equation}
where $\varphi(\tau)$ is a slowly varying phase factor which depends on the
properties of the cantilever. The damping of the coherent oscillations is given
by
\begin{eqnarray*}
\Gamma(\tau)&=&\frac{4\lambda^2(2{\bar N}+1)}{\hbar^2(\omega_{\rm m}^2+\gamma^2)}
\left\{
\gamma\tau
-\frac{2\gamma\omega_{\rm m}}{\gamma^2+\omega_{\rm m}^2}
{\rm e}^{-\gamma \tau}\sin(\omega_{\rm m}\tau)\right.\\
&&\left. +\left(\frac{\gamma^2-\omega_{\rm m}^2}{\gamma^2+\omega_{\rm m}^2}
\right)\left[{\rm e}^{-\gamma \tau}\cos(\omega_{\rm m}\tau)-1\right]
\right\}.
\end{eqnarray*}
Again we define the envelope of
${\rm P}_{\rm th}(|-\rangle)$  by setting the total phase in the 
square brackets of Eq.\ (\ref{two}) to zero.

\begin{figure}[t]
\mbox{\epsfig{file=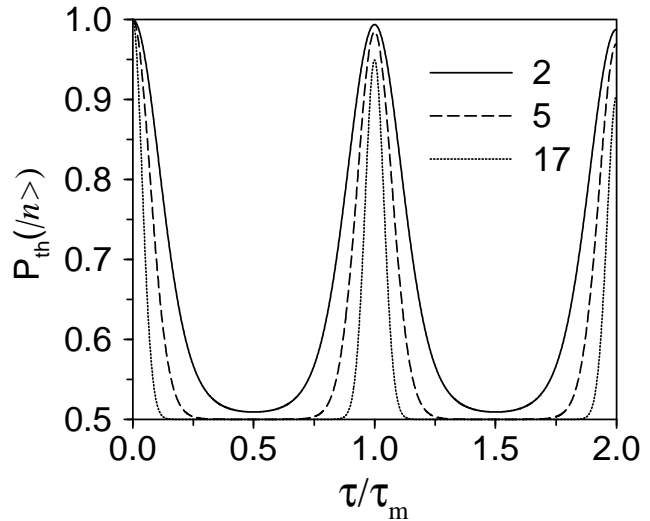, width=3.4in}}
\caption{Envelope of ${\rm P_{th}}(|-\rangle)$, 
including the cantilever's environment, as a function
of wait time for $Q=1000$. The figures in the legend correspond to the
values of the quantity $4\kappa^2
(2{\bar N}+1)$.}
\end{figure}

The energy damping rate in the model, $2\gamma$, can be estimated
empirically by measuring the quality factor of the cantilever, $Q$, 
since $2\gamma=\omega_{\rm m}/Q$. Fig.~3 shows the envelope of 
${\rm P}_{\rm th}(|-\rangle)$ when the coupling of the cantilever to the
environment is included, for  $Q=1000$ as a function of the quantity 
$(2\lambda/\hbar\omega_{\rm m})^2(2{\bar N}+1)$. The series of curves 
shown could be obtained, for example, by setting the temperature at 
$30~{\rm mK}$ and sweeping the coupling strength $\kappa$ from $0.14$ 
to $0.41$. In the
presence of a finite damping rate, the recoherences are indeed 
suppressed progressively as either the temperature or the cantilever--Cooper
box coupling is increased. Notice that because of the predicted dependence
of the decoherence rate of the cantilever on the coupling and temperature, 
it would be possible to separate out the effect of the cantilever's environment
from  other contributions causing decoherence of the Cooper box.

The final stage in the process is to read out the charge state of the
Cooper box using the rf-SET. At the end of the control stage, the rf-SET is 
tuned away from 
the Coulomb blockade region and a non-zero drain-source voltage 
applied, resulting in a tunneling current through the SET. As a 
result of the capacitive coupling $C_{\rm int}$ between the Cooper box and 
SET, the SET island voltage  will be affected by the 
Cooper box island charge. Hence, the SET tunneling  
current probes the Cooper box charge state. If the lifetime of the 
Cooper box state is determined 
by the rf-SET island voltage and quantum electromagnetic mode 
fluctuations acting back on the box, then the condition for the measurement time 
to be shorter than this lifetime 
is~\cite{devoret,aassime,makhlin}:
\[
\frac{\tau_{\rm decay}}{\tau_{\rm measure}}=\left(\frac{4E_{\rm 
C}}{E_{\rm J}}\right)^{2}\frac{\hbar^{2}}{S_{\rm V}(\delta q)^{2}}>1,
\]
where $\delta q$ is the charge sensitivity of the rf-SET and
$S_{\rm V}$ is the sum of the SET island and electromagnetic mode 
voltage noise evaluated at the 
Cooper state oscillation frequency $\omega=4E_{\rm C}/\hbar$. 
Using the values $E_{\rm C}=150~\mu{\rm eV}$ and $E_{\rm J}=4~\mu{\rm eV}$, 
resulting from the above state control condition, the 
electromagnetic-mode dominated voltage noise
$S_{\rm V}=0.14~{\rm nV}^{2}/{\rm Hz}$ at $4E_{\rm C}/h=145~{\rm GHz}$, and the 
 value for the rf-SET charge sensitivity $\delta q=6.3~\mu e/\sqrt{\rm Hz}$
determined experimentally in ref.~\onlinecite{aassime}, 
we have $\tau_{\rm decay}/\tau_{\rm measure}=1.7\times 10^3$. 
Choosing, for example, 
$C_{\rm int}/C_{\rm J}=0.1$, the respective times are in fact  
$\tau_{\rm measure}=4~{\rm ns}$ and $\tau_{\rm decay}=7~\mu{\rm s}$. 
However, the actual lifetime 
is likely to be somewhat smaller than $7~\mu{\rm s}$,  
limited by Cooper box offset charge noise~\cite{nakamura2,cottet}, but  
it certainly exceeds $4~{\rm ns}$~\cite{nakamura1}. 
Thus, provided $C_{\rm int}$ is not too small, it should be 
possible read out the charge state.             
  
The scheme we have detailed 
provides a feasible way of probing the quantum coherence of a micromechanical 
system. The calculation of the decoherence rate of the cantilever due to 
interactions with an oscillator-bath predicts that mechanical systems of the 
kind now available  could display quantum coherence over time-scales of a few 
periods. The analysis detailed here
is readily adapted to include more elaborate pulse sequences, such as that
used by Nakamura {\it et al}.~\cite{nakamura2} to reduce the intrinsic
decoherence time of the Cooper box, without significantly 
affecting the conclusions.    

We thank R. Lifshitz and A. MacKinnon for 
useful discussions. This work was supported in part by 
the National Security Agency (NSA), the  Advanced Research and 
Development Activity (ARDA), the  Army Research Office (ARO), and
the Engineering and Physical Sciences Research Council (EPSRC).

\end{document}